\documentclass[aps,prb,twocolumn,superscriptaddress]{revtex4}
\usepackage{ae}
\usepackage[T1]{fontenc}
\usepackage[ansinew]{inputenc}
\usepackage{amsmath}
\usepackage{amssymb}
\usepackage{graphicx}
\usepackage{color}
\usepackage[colorlinks]{hyperref}
\usepackage{epstopdf}

\def\sig{{\mbox{\boldmath{$\sigma$}}}}

\def\sig{{\mbox{\boldmath{$\sigma$}}}}
\usepackage{soul,xcolor}
\setstcolor{blue}                   

\def\sig{{\mbox{\boldmath{$\sigma$}}}}

\def\pf{{\mbox{\boldmath{$\varphi$}}}}

\begin{document}

\date{\today}

\title{DC spin generation by  junctions with AC driven spin-orbit interaction}

\author{M. Jonson}
\affiliation{Department of Physics, University of Gothenburg, SE-41296 G\"{o}teborg, Sweden}
\author{R. I. Shekhter}
\affiliation{Department of Physics, University of Gothenburg, SE-41296 G\"{o}teborg, Sweden}
\author{O. Entin-Wohlman}
\affiliation{Raymond and Beverly Sackler School of Physics and Astronomy, Tel Aviv University, Tel Aviv 69978, Israel}
\author{A. Aharony}
\affiliation{Raymond and Beverly Sackler School of Physics and Astronomy, Tel Aviv University, Tel Aviv 69978, Israel}
\author{H. C. Park}
\affiliation{Center for Theoretical Physics of Complex Systems, Institute for Basic Science (IBS), Daejeon 34051, Republic of Korea}
\author{D. Radi\'{c}}
\affiliation{Department of Physics, Faculty of Science, University of Zagreb, Bijeni\v{c}ka 32, Zagreb 10000, Croatia}

\begin{abstract}
An  unbiased one-dimensional weak link between two terminals, subjected to the Rashba spin-orbit interaction caused by an AC electric field which rotates periodically in the plane perpendicular to the link,
is shown to inject spin-polarized electrons into the  terminals. The injected spin-polarization has  a DC component along the link and a rotating transverse component in the perpendicular plane. In the  low rotation-frequency regime, these polarization components are proportional to the frequency.
The DC component  of the polarization vanishes for a linearly-polarized electric field.

\end{abstract}

\maketitle

\section{Introduction}
\label{SecI}

Spintronics takes advantage of the electronic spins in designing a variety of applications, including  giant magnetoresistance sensing, quantum computing,  and quantum-information processing. \cite{wolf,zutic,Kim} A promising approach for the latter  exploits mobile qubits, which carry the quantum information via the spin polarization of moving electrons. The spins of mobile electrons can be manipulated  by  the spin-orbit interaction (SOI), which  causes  the spin of an electron moving through a spin-orbit active material (e.g.,  semiconductor heterostructure \cite{Kohda})  to rotate around an effective magnetic field. \cite{winkler,manchon} In the particular case of the Rashba SOI, \cite{rashba} the magnitude and direction of this field can be tuned by  gate voltages. \cite{Nitta,Sato,Beukman,comDres}
The Rashba SOI is mostly significant at surfaces and interfaces because of  strong internal uncompensated atomic electric fields perpendicular to the surface/interface. These occur since the (weaker)  surface/interface potential breaks the symmetry of the atomic orbitals there, so that the corresponding  strong atomic fields no longer cancel as they do in the bulk. An electric field induced by external gates can then modulate the resulting SOI to a certain extent  by changing the degree of orbital asymmetry.


One aim of spintronics is to  build logic devices,  \cite{Kim} which produce spin-polarized electrons, so that one can use their electronic spinors as qubits. In the simplest device, electrons move between two large electronic reservoirs, via a nano-scale quantum network. 
For this two-terminal case, the time-independent SOI that obeys time-reversal symmetry cannot generate spin splitting. \cite{bardarson} Time-reversal symmetry can be broken  by applying  a magnetic field, either via a magnetic flux, which penetrates SOI-active loops of Aharonov-Bohm interferometers,  \cite{lyanda,us,Saarikoski} or by a Zeeman  magnetic  field.
\cite{Shmakov,Nagasawa} Alternatives utilize 
ferromagnetic terminals. \cite{datta,sarkar} 
%

Here we explore yet another means to break time-reversal symmetry, exploiting time-dependent Hamiltonians.  Several papers proposed the generation  of spin-splitting by  quantum spin pumping, in which  different terms in the system's Hamiltonian vary slowly periodically with time. Some of these require DC or AC magnetic fields. \cite{marcus,wang,janine} Here we concentrate on all-electrical devices, which pump polarized electrons. One such  device used  an out-of-phase oscillation of the  heights of the barriers representing the contacts between a planar quantum dot and the two leads  to yield a spin current with polarization perpendicular to the plane. \cite{brouwer} Alternatively, polarized spins  were created  by  periodic variations of one barrier height and of the strength of a uniaxial-SOI  (induced by an electric field perpendicular to the  quantum dot's plane). \cite{governale} In a third example, a one-dimensional wire was split into two regions, with two differently-oriented SOI-generating electric fields which oscillate periodically with time. \cite{avishai}
In  these examples, the two gate voltages act at different locations of the system, and the calculation yields only the average spin current, integrated over a period of the oscillation.

\begin{figure}
\includegraphics[width=\columnwidth]{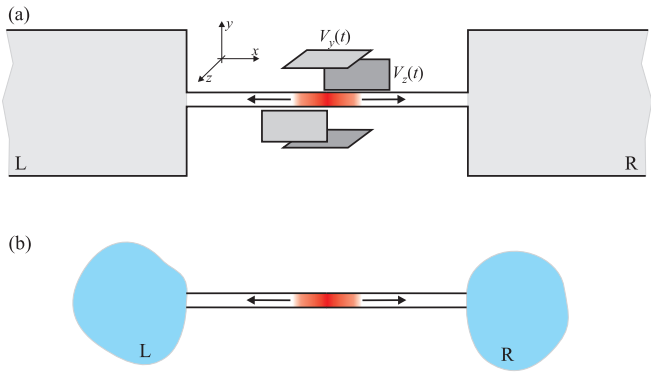}
\caption{Schematic visualizations of  devices proposed in the text. (a)
A  spin-orbit-active weak link connects two contacts, $L$ and $R$, to form a closed circuit. \cite{circuits} The time-dependent spin-orbit interaction is generated by two perpendicular gates, whose potentials $V_{y}(t)$ and $V_{z}(t)$ oscillate slowly in time with frequency $\Omega$.  The arrows within the weak link indicate the directions in  which polarized electron-spins are flowing.  (b) An open-circuit version of (a) where spin is accumulated in two terminals leading to a magnetization that can be measured.
}
\label{Fig1}
\end{figure}

Below we consider
the possibility to activate spin splitting via weak links (also called `junctions') by breaking time-reversal symmetry with  an AC Rashba SOI created by an electric field that rotates slowly with frequency $\Omega$
perpendicularly to the (one-dimensional) weak link.
A rotating field can result from two external fields along perpendicular directions,  which are normal to a thin cylindrical wire. When the two fields oscillate periodically with time, with a phase difference of $\pi/2$, the resultant vector rotates around the wire. Such fields can be produced by gate voltages $V_y(t)$ and $V_z(t)$, applied to  electrodes  as in
\ref{Fig1}(a). \cite{com2}
They can  also be generated by rotating
a bent wire  periodically under a uniform electric field,   \cite{extension}  or  from a circularly-polarized electromagnetic field.
Even in absence of a bias we find that  a {\em time-independent} DC flow, towards both terminals, of electrons whose spins are polarized  parallel to the junction's direction,  is created  in the junction  [as indicated  by arrows in the weak link shown in Fig.~\ref{Fig1}]. In addition, the time-dependence of the SOI in the weak link gives rise to transverse components of  the polarization, which rotate in the plane perpendicular to the junction  in parallel to the effective SOI magnetic field. These transverse components vanish upon averaging over a period, and thus  would not appear in the `standard' spin-pumping approach. We analyze in detail the case of a circularly-polarized electric field, and then extend the discussion to allow for an elliptic variation of the field,  all the way to the limit of a longitudinal uniaxial oscillation, where the
DC spin polarization is found to vanish.

Our model is described in Sec. \ref{SecII}, where we also give general expressions for the charge and spin currents. The detailed derivation of these currents is presented in Appendix \ref{AppA}. Sections \ref{SecIII} and \ref{SecIV} then present explicit results for these currents, for a circularly  and an elliptically rotating electric field. Technical details of these calculations appear in Appendix \ref{AppB}. Our conclusions are then discussed in Sec. \ref{SecV}.

\section{Model and currents}
\label{SecII}

Electronic transport
through a spin-orbit-active weak link
can be analysed within the framework of an effective tunneling Hamiltonian,
\begin{align}
\label{Htun}
{\cal H}^{}_{\rm tun}(t)=\sum_{k,p}\sum_{\sigma,\sigma'}\Big ([W^{}_{LR}(t)]^{}_{\sigma\sigma'}c^{\dagger}_{k\sigma}c^{}_{p\sigma'}+{\rm H.c.}\Big )\ ,
\end{align}
where
$c^{\dagger}_{k\sigma}$ ($c^{}_{k\sigma}$)
creates (annihilates)
an electron of wave vector $k$ and spin index $\sigma$ in the left electrode; \cite{com1} the wave vectors on the right  are denoted by $p$.
The tunneling amplitude $W_{LR}(t)$ from right to left is  a (2$\times$2) matrix in spin space,   independent of the wave vectors, i.e.,  it is approximated by its value at the Fermi energy.
For the Rashba SOI, this tunneling amplitude has the form
\begin{align}
W^{}_{LR}(t)=W^{}_0 \exp[i\pf^{}_{\rm AC}(t)]\ ,
\label{WT}
\end{align}
where $W_{0}$  sets the magnitude of the tunnel coupling (in units of energy) and the Aharonov-Casher \cite{Aharonov-Casher,ora}
phase  operator is
\begin{align}
\pf^{}_{\rm AC}(t)=k^{}_{\rm so}d [\hat{\bf x}\times\hat{\bf n}(t)]\cdot\sig\ .
\label{PAC}
\end{align}
Here,
$\sig=(\sigma_{x},\sigma_{y},\sigma_{z})$ is the vector of the Pauli matrices,
the unit vector $\hat{\bf x}$  is along the weak link whose length is $d$,
and $k_{\rm so}$ is the strength of the SOI (in units of inverse length), resulting from an electric field directed along  $\hat{\bf n}(t)$; its explicit form is specified below.

Equation (\ref{PAC}) was derived in Ref. \onlinecite{PRL2016} for a time-independent SOI, $\hat{\bf n}(t)=\hat{\bf n}$.
A generalization to the time-dependent case considered here would in principle require an analysis of electron tunneling through a device, which includes the non-trivial dynamics of the SOI. One would expect such an analysis to lead to a non-local temporal relation between the Aharonov-Casher phase and the electric field direction. In the present paper we divide the calculation into two steps: in the first step, we neglect this non-trivial dynamics, and treat the time $t$ in Eq. (\ref{PAC}) as a parameter. In particular, we expect this approximation to be valid when the time-dependence of the SOI is slower than all other time scales in the system. Assuming that the electric field rotates with frequency $\Omega$, the result may be justified for $\Omega\tau\ll 1$, where $\tau$ is the electron dwell time before the electron escapes from  the weak link to an adjacent reservoir (to be estimated below). In the second step we use a full time-dependent perturbation expansion in the tunneling matrix elements $W^{}_{LR}(t)$.

The entire tunnel junction is modelled by the Hamiltonian
\begin{align}
{\cal H}={\cal H}^{}_{\rm leads}+{\cal H}^{}_{\rm tun}(t)\ ,
\label{H}
\end{align}
where the first term describes the non-polarized (free-electron) leads
\begin{align}
\label{Hleads}
{\cal H}^{}_{\rm leads}=\sum_{k ,\sigma}\epsilon^{}_{k}c^{\dagger}_{k\sigma}c^{}_{k\sigma}+
\sum_{p ,\sigma}\epsilon^{}_{p}c^{\dagger}_{p\sigma}c^{}_{p\sigma}\ ,
\end{align}
and $\epsilon_{k(p)}$ is the single-electron energy in the left (right) lead.

The rates of change of the particle number and of the magnetization in the left lead are determined by the rate matrix
\begin{align}
\label{RRR}
[R^{}_{L}]^{}_{\sigma\sigma'}(t)\equiv\frac{d}{dt}\sum_{k}\langle c^{\dagger}_{k\sigma}(t)c^{}_{k\sigma'}(t)\rangle\ ,
\end{align}
where angular brackets indicate quantum averaging.
Specifically, the particle
current into the left electrode, $I^{}_{L}(t)$,  is  the rate of change of the particle occupation there,
\begin{align}
I^{}_{L}(t)=\frac{d}{dt}\sum_{\sigma}\sum_{k}\langle c^{\dagger}_{k\sigma}(t)c^{}_{k\sigma}(t)\rangle\equiv \sum_{\sigma}[R^{}_{L}(t)]^{}_{\sigma\sigma}\ ,
\label{opIL}
\end{align}
while the rate of change of the total spin
in the left terminal, and hence the spin current into that terminal, is $(\hbar/2)\dot{\bf M}^{}_{L}$, with
\begin{align}
\dot{\bf M}^{}_{L}=
\frac{d}{dt}\sum_{\sigma,\sigma'}\sum_{k}\langle c^{\dagger}_{k\sigma}(t)\sig^{}_{\sigma\sigma'}c^{}_{k\sigma'}(t)\rangle\equiv\sum_{\sigma,\sigma'}[R^{}_{}(t)]^{}_{\sigma\sigma'}\sig^{}_{\sigma\sigma'}\ .
\label{opML}
\end{align}
It follows that the rate of change of magnetization in the left lead is $(g\mu^{}_B/2) \dot{\bf M}^{}_{L}$ ($g$ is the $g-$factor of the electrodes, and $\mu_{\rm B}$ is the Bohr magneton, see Sec. \ref{SecV}).

As detailed in Appendix A [see Eq. (\ref{R})], perturbation theory to second order in the tunneling amplitude gives ($\hbar = 1$, $\eta\rightarrow 0^{+}$)
 \begin{align}
&[R^{}_{L}]^{}_{\sigma\sigma'}(t)=\sum_{k,p}\Big [f^{}_{R}(\epsilon^{}_{p})-f^{}_{L}(\epsilon^{}_{k})\Big ]\int_{-\infty}^{t}dt^{}_{1}e^{\eta t^{}_{1}}\nonumber\\
&\times\Big (e^{i(\epsilon^{}_{k}-\epsilon^{}_{p})(t-t^{}_{1})}[W^{}_{LR}(t)W^{\dagger}_{LR}(t^{}_{1})]^{}_{\sigma'\sigma}+{\rm H.c.}\Big )\ .
\label{Ra}
\end{align}
Here,
\begin{align}
f^{}_{L(R)}(\epsilon^{}_{k(p)})=\big(\exp[(\epsilon^{}_{k(p)}-\mu^{}_{L(R)})/(k^{}_BT)]+1\big)^{-1}
\label{fermi}
\end{align}
 is the equilibrium Fermi function in the left (right) lead, whose  chemical potential is $\mu^{}_{L(R)}$.


\section{Circularly rotating field}
\label{SecIII}

For a circularly-polarized electric field
${\bf n}(t)=\cos(\Omega t)\hat{\bf z}-\sin(\Omega t)\hat{\bf y}$, and thus the tunneling amplitude is
\begin{align}
W^{}_{LR}(t)=W^{}_{0}[\cos(k^{}_{\rm so}d)+i\sin(k^{}_{\rm so}d)\sig\cdot\hat{\bf v}(t)]\ ,
\label{W}
\end{align}
where
\begin{align}
\hat{\bf v}(t)=[0,\cos(\Omega t),\sin(\Omega t)]
\label{v}
\end{align}
lies in  $y-z$ plane as well.
Consequently,
\begin{widetext}
\begin{align}
&W^{}_{LR}(t)W^{\dagger}_{LR}(t^{}_{1})/|W^{}_{0}|^{2}=\cos^{2}_{}(k^{}_{\rm so}d)+\sin^{2}_{}(k^{}_{\rm so}d)\hat{\bf v}(t)\cdot\hat{\bf v}(t^{}_{1})\nonumber\\
&+i\sig\cdot\Big (\cos(k^{}_{\rm so}d)\sin(k^{}_{\rm so}d)[\hat{\bf v}(t)-\hat{\bf v}(t^{}_{1})]+
\sin^{2}_{}(k^{}_{\rm so}d)\hat{\bf v}(t)\times\hat{\bf v}(t^{}_{1})\Big )\ .
\label{pro}
\end{align}
\end{widetext}

The particle current, Eq. (\ref{opIL}), requires the trace of the matrix $R^{}_L(t)$, hence [see Eq. (\ref{Ra})] the trace of $W^{}_{LR}(t)W^\dagger_{LR}(t^{}_1)$. As shown in Appendix \ref{AppB}, this trace depends only on $(t-t^{}_1)$. Assuming also
that the densities of states in the terminals are energy independent,  \cite{mjcomment}
${\cal N}_{L(R)}(\epsilon)={\cal N}_{L(R)}(\epsilon_{\rm F})\equiv {\cal N}_{L(R)}$ (wide-band approximation), and if $\mu_R - \mu_L = eV$, then in the limit of low bias-voltage $V$ and low temperature Eq. (\ref{IL1}) yields
\begin{align}
\label{ILresult}
I^{}_L = G V/e; \quad G=4\pi^2 \vert W_0  \vert^2 {\cal N}^{}_{L}{\cal N}^{}_{R} G^{}_0\ ,
\end{align}
where $G_0=e^2/(\pi \hbar)$ is the quantum of conductance. Clearly, the particle current is not affected by the spin-orbit interaction.

The current into the right terminal,
$I^{}_R$,   
is 
(see Appendix B) $I_{R}=-I_{L}$,
demonstrating that particle number is conserved in the junction.
For the spin currents,  however, there is no such conservation law, and  in fact spin-flip transitions generated by the
SOI in the weak link may result in the accumulation of spin polarization.
Indeed, as seen in Eqs.~(\ref{MLx}) and (\ref{MLtr}) below for the spin polarization in the left lead,  interchanging $L$ with $R$ in each of them to obtain the spin polarization in the right one leaves them intact, $\dot{\bf M}_{L}(t)=\dot{\bf M}_{R}(t)$;
the total spin is not conserved, and the junction injects the same amount of spin polarization into the two leads, even in the absence of any bias voltage.

From Eqs. (\ref{opML}) and (\ref{Ra}), the spin current requires the trace of $\sig W^{}_{LR}(t)W^\dagger_{LR}(t^{}_{1})$.
Equation  (\ref{dMx}) then implies
that (with the same assumptions as above) the $x-$component of the spin-polarization flow is
proportional to
 \begin{align}
&\dot{ M}^{}_{L,x}=(G/G^{}_0){\cal F}(\Omega)
\sin^{2}_{}(k^{}_{\rm so}d)\ ,
\label{MLx}
\end{align}
where
\begin{align}
{\cal F}(\Omega )&=\int \frac{d\omega d\omega '}{2\pi}[f^{}_{L}(\omega)-f^{}_{R}(\omega')]\nonumber\\
&\times[\delta(\omega-\omega '+\Omega)-
\delta(\omega-\omega '-\Omega)]\ .
\label{F}
\end{align}
Interestingly, the $x-$component of the injected spin polarization is time-independent; the AC electric field yields a DC polarization in the leads, parallel to the junction.
At small $\Omega$, the difference of the two delta-functions in Eq.~(\ref{F}) is proportional to $\Omega$,
indicating  inelastic processes: electrons  exchange  photons of energy $\Omega$ with the electric field, and accordingly flip their spins.   For instance, at zero temperature only absorption processes  are allowed in an un-biased junction
; these
lead to pumping of the $x-$component of the spin polarization into the terminals.

The transverse components of the spin-polarization flow do oscillate with time, since (see Appendix B) 
\begin{align}
\dot{\bf M}^{\rm tr}_{L}(t)=\frac{G}{G^{}_0}\frac{{\cal F}(\Omega)}{2}\sin (2k^{}_{\rm so}d)
[0,\sin(\Omega t),-\cos(\Omega t)]\ .
\label{MLtr}
\end{align}
The sum of the two transverse spin components is directed along  the vector $[0,\sin(\Omega t),-\cos(\Omega t)]$.
Integration over time yields  a transverse spin polarization ${\bf M}^{\rm tr}_{L}(t) = \int^t\dot{\bf M}^{\rm tr}_{L}(t^\prime)dt^\prime$, which is parallel to the effective  magnetic field, i.e., to $\hat{\bf v}(t)$,  Eq.~(\ref{v}). 



\section{Elliptically rotating field}
\label{SecIV}

The DC character of the flow of the longitudinal  ($x-$) component of the spin polarization is our main result. It is a remarkable consequence of the AC electric field responsible for the SOI and crucially depends on the fact that this electric field is rotating in the plane perpendicular to the weak link.
To elucidate this point we allow for  different amplitudes of the electric field components oscillating in the two transverse ($y-$ and $z-$) directions.
 In that case the tunneling amplitude takes the form
\begin{align}
W_{LR}(t)/W^{}_{0}=\cos[U(t)k^{}_{\rm so}d]+i\sin[U(t)k^{}_{\rm so}d]\sig\cdot\hat{\bf u}^{}_{}(t)\ ,
\label{Wn}
\end{align}
where
\begin{align}
\hat{\bf u}_{}(t)={\bf U}(t)/U(t)\equiv
[0,\cos(\Omega t),\gamma\sin(\Omega t)]/U(t)\ ,
\end{align}
and
\begin{align}
 U(t)=\sqrt{\cos^{2}_{}(\Omega t)+\gamma^{2}_{}\sin^{2}_{}(\Omega t)}\ .
 \label{Rt}
 \end{align}
As seen,  $\gamma$
 measures the deviation away from the circular polarization: $\gamma=0$ corresponds to a linear-polarized electric field, while $\gamma=1$ restores the circularly-polarized field, Eq.~(\ref{v}).

 Whereas a circularly-polarized electric field  implies single-photon absorption and emission processes [as expressed by the delta-functions in Eq.~(\ref{F})], the intricate time dependence [see Eq.~(\ref{Rt})] of the non-circular polarization leads to an infinite Fourier series in powers of $\exp[i n\Omega t]$, and consequently to an infinite series of delta-functions expressing multiple-photon processes of emission and absorption. However, upon retaining only terms up to second order in $(k_{\rm so}d)$ Eq.~(\ref{Wn}) becomes
\begin{align}
W_{LR}(t)/W^{}_{0}\approx 1-[U(t)(k^{}_{\rm so}d)]^2/2+i(k^{}_{\rm so}d)\, \sig\cdot{\bf U}^{}_{}(t)\ .
\label{Wn1}
\end{align}

 The spin part here, which determines $\dot{\bf M}^{}_L(t)$, has the same form as in a similarly-expanded Eq.~(\ref{W}), except that the coefficient of $\sigma^{}_z$ is multiplied by $\gamma$. Repeating the previous calculation, this  reproduces Eqs.~(\ref{MLx}) and (\ref{MLtr}), except that the $x-$ and  $z-$components of $\dot{\bf M}^{}_L$ are now multiplied by $\gamma$. Therefore, in the longitudinal limit $\gamma=0$ and for a small SOI there is no DC spin current, while the transverse spin current oscillates only in the $\hat{\bf y}-$direction.

\section{Discussion}
\label{SecV}

An external magnetic field, whose orientation varies with time, prevents the spin of an electron from being a good quantum number and induces spin-flip transitions. Spin pumping of electrons in semiconductors by circularly-polarized light \cite{astakhov} and spin currents generated by magnetization dynamics in conducting ferromagnets \cite{brataas} are well known consequences of this fact. These phenomena are, however, distinct from the effect discussed here. This is because a time-independent SOI, in contrast to a Zeemann coupling, preserves time reversal symmetry and {\it hence \cite{bardarson} does} not affect electronic transport properties. Rather, a Rashba type SOI generates an energy-independent Aharonov-Casher phase, which can only affect electron transport if time reversal symmetry is broken in some other way. In our work this is achieved by assuming that the SOI is generated by a time-dependent (AC) electric field. If a constant-magnitude electric field rotates in the plane perpendicular to a one-dimensional wire the DC spin current (spin polarization flow) given by Eq. (\ref{MLx}) is generated.

To elaborate on the physical reason for the generation of this DC spin current we note that semi-classically the effect of the SOI (here restricted to act in the weak link only) can be viewed as a precession of the spin of the electrons during their passage through the link. When the SOI is due to an electric field that rotates in the plane perpendicular to the weak link, the direction of the spin-rotation axis, which is perpendicular to both the electric field and the direction of electron propagation, rotates in the same plane and there is no spin-component that can be chosen as an integral of motion. Since  spin is not conserved in the weak link, excess spin is accumulated there. This is why the electron flow injected from the weak link into  the left as well as the right lead carries net spin corresponding to a spin current through the leads generated in the weak link. A remarkable result of our calculation is that the total rate of spin generation in the link (the number of spins injected into the leads per unit time) does not depend on time if the electric field responsible for the SOI has a constant magnitude and rotates by a constant frequency.

In reality, the flow of the spin-polarized electrons
injected from the junction into the adjacent parts of the terminals has a certain spatial dependence.
For one-dimensional leads (
in the absence of the SOI and magnetic fields), we expect the extra charge and spin polarization in the terminals to follow a classical trajectory  with the Fermi velocity $v^{}_{\rm F}$ in the leads, e.g.
$M^{}_{L,x}(r,t)=M^{}_{L,x}(0,t-r/v^{}_{\rm F})$ at a distance $r$ from the edge of the junction, up to a certain length determined by the  impurity scattering length in the terminal.
The periodic rotation of the transverse spin components will translate into a periodic rotation in space.
In higher dimensions, the ballistic electronic motion can be treated as in the theory of point-contact spectroscopy of metals, with the corresponding densities decaying as $(r^{}_0/r)^\xi$,
where $r^{}_0$ characterizes  the cross-section of the junction and $\xi = 2\  (\xi = 1)$
in the ballistic (diffusive) transport regime.  \cite{30}

In the closed-circuit configuration\cite{circuits} sketched in Fig.~\ref{Fig1}(a) the  magnetization injected into the leads can be measured, e.g., by a properly positioned SQUID,  
or by
a magnetic-resonance force microscope.
Alternatively, an open circuit\cite{circuits}  such as the one sketched in Fig.~\ref{Fig1}(b), where magnetization is accumulated in two terminals, can be used. Here low-dimensional contacts connect the weak link to terminals whose linear dimension significantly exceeds the cross section of the contacts so that the terminals can be thought of as reservoirs where injected  polarized spins spend a significant time --- much longer than the spin relaxation time --- before they are reflected back to the weak link.

An estimate for the amount of magnetization accumulated  in one of the terminals during a time interval of the order of the spin relaxation time $\tau^{}_{s}$   will then be $(g\mu_B/2)\dot{M}_{L,x}\tau^{}_{s}$. Using Eqs.~(\ref{MLx}) and (\ref{F}) 
without any bias voltage, $\mu_L = \mu_R$, we find that in the small $\Omega$ limit
\begin{align}
\dot{M}^{}_{L,x}=(\Omega/\pi) (G/G^{}_0)\sin^{2}(k^{}_{\rm so} d)\ .
\end{align}
Consider now an SOI-active weak link in the form of an InAs nanowire 
and adopt the value $k_{\rm so}=1/(100 \,{\rm nm})$ measured by
Scher\"{u}bl {\it et al.}  \cite{nygard}
A wire length of $d=100$nm would then give 
$k_{\rm so}d=1$ and hence $\sin^{2}(k^{}_{\rm so}d)\approx 1$.
For $\Omega=2\pi \times 20$ GHz 
and using the typical value $G \sim 0.5 G_0$ for the normal conductance of InAs nanowires
\cite{nygard, chuang, dayeh}
one then finds that $\dot{M}^{}_{L,x}\approx 2 \times 10^{10}$ s$^{-1}_{}$. 
Next consider 
$n$-type bulk GaAs terminals [see Fig.~\ref{Fig1}(b)] moderately doped 
to give a low electron density of $1 \times 10^{16}\,$ cm$^{-3}$, for which spin relaxation times as long as $\tau_{s}=100\,$ns have been measured at low temperatures. \cite{kikkawa}
Using the measured value $g=-0.45$ for the $g$-factor of conduction electrons in bulk GaAs,  \cite{pidgeon} we then arrive at the conclusion that spins corresponding to a magnetization of about 500 Bohr magnetons may accumulate in the terminals, which if they were cubes with side-lengths of 1$\,\mu$m would contain $\sim$10,000 electrons.


As stated, we expect our approximation (\ref{PAC}) to be valid when
 $\Omega \ll 1/\tau$, or equivalently when $\hbar \Omega \ll \Gamma$, where $\Gamma=\hbar/\tau$ is the level width in the wire. For this purpose we use the estimate $\Gamma = D \hbar v_{\rm F}/d$, where $D \sim G/G_0 = 0.5$ is the transparency of the barriers between the wire and the terminals
and $v_{\rm F} = \ell_e e/(m^\ast \mu_e)$ is the Fermi velocity of electrons in the wire, here related to the electron mobility $\mu_e$, the electron mean free path $\ell_e$, and the  effective mass $m^\ast =0.023 m$. Using the typical InAs-nanowire values $\mu_e = 3000\,$ cm$^2$/(Vs) and $\ell_e = 55\,$nm (taken from Ref.~\onlinecite{dayeh}) we find that $v_{\rm F}\approx 1 \times 10^8\,$cm/s and hence $\Gamma \approx 3\,$meV. Since $\hbar \Omega \approx 0.1\,$meV for $\Omega = 2\pi \times 20\,$GHz we are indeed in the low-frequency regime.

In conclusion we have shown that a rotating electric field, acting on a weak link between two non-magnetic metals, generates both a DC spin current and transverse spin components which rotate around the link,  flowing into both metals. This is a novel simple device, with potential uses as a logic element in quantum data processing. Our estimates show that it can realistically be made with existing materials and technology.

\begin{acknowledgments}

We thank Jungho Suh and Chulki Kim for fruitful discussions.
This research was  partially supported by the Israel Science Foundation (ISF), by the infrastructure program of Israel Ministry of Science and Technology under contract 3-11173,  by the Pazy Foundation, by
the Croatian Science Foundation, project IP-2016-06-2289, and by the
QuantiXLie Centre of Excellence, a project cofinanced by the Croatian
Government and the European Union through the European Regional
Development Fund - the Competitiveness and Cohesion Operational Programme
(Grant KK.01.1.1.01.0004). MJ, RJS, OEW, AA, and DR acknowledge the hospitality of the PCS at IBS, Daejeon, Korea, where part of this work was done.
\end{acknowledgments}

\appendix


\section{Derivation of particle- and spin currents  in the model junction}
\label{AppA}

As explained in the main text,  the tunnel junction is modelled by the Hamiltonian
${\cal H}={\cal H}^{}_{\rm leads}+{\cal H}^{}_{\rm tun}(t)$, Eq.~(\ref{H}),
in which ${\cal H}^{}_{\rm leads}$, given by Eq. (\ref{Hleads}), describes the non-polarized (free-electron) leads.
Tunneling between the leads is described by the time- and spin-dependent
Hamiltonian (\ref{Htun}).

For non-polarized leads, one  finds
\begin{align}
\label{opR}
[R^{}_{L}]^{}_{\sigma\sigma'}(t)&\equiv\frac{d}{dt}\sum_{k}\langle c^{\dagger}_{k\sigma}(t)c^{}_{k\sigma'}(t)\rangle\\
&=i\sum_{k,p}\sum_{\sigma^{}_{1}}\Big ([W^{\ast}_{LR}(t)]^{}_{\sigma\sigma^{}_{1}}\langle c^{\dagger}_{p\sigma^{}_{1}}(t)c^{}_{k\sigma'}(t)\rangle
\nonumber\\
& \hspace{20mm}- [W^{}_{LR}(t)]^{}_{\sigma'\sigma^{}_{1}}\langle c^{\dagger}_{k\sigma}(t)c^{}_{p\sigma^{}_{1}}(t)\rangle\Big)\ .\nonumber
\end{align}
The quantum averages in the first and second terms on the right hand-side of Eq. (\ref{opR}) are
calculated to lowest order in the tunneling (using units in which $\hbar=1$)
\begin{align}
\label{pt}
\langle c^{\dagger}_{p\sigma^{}_{1}}(t)c^{}_{k\sigma'}(t)\rangle&=i\int_{-\infty}^{t}dt^{}_{1}e^{\eta t^{}_{1}}e^{i(\epsilon^{}_{p}-\epsilon^{}_{k})(t-t^{}_{1})}\\
&\hspace{10mm}\times [W^{}_{LR}(t^{}_{1})]^{}_{\sigma'\sigma^{}_{1}}[f^{}_{L}(\epsilon^{}_{k})-f^{}_{R}(\epsilon^{}_{p})]\ ,\nonumber\\
\langle c^{\dagger}_{k\sigma}(t)c^{}_{p\sigma^{}_{1}}(t)\rangle&=i\int_{-\infty}^{t}dt^{}_{1}e^{\eta t^{}_{1}}e^{i(\epsilon^{}_{k}-\epsilon^{}_{p})(t-t^{}_{1})}\nonumber\\&\hspace{10mm}\times [W^{\ast}_{LR}(t^{}_{1})]^{}_{\sigma\sigma^{}_{1}}[f^{}_{R}(\epsilon^{}_{p})-f^{}_{L}(\epsilon^{}_{k})]\ ,
\nonumber
\end{align}
where $\eta\rightarrow 0^{+}$.
Inserting Eqs. (\ref{pt}) into Eq. (\ref{opR}) gives
\begin{align}
\label{R}
[R^{}_{L}]^{}_{\sigma\sigma'}(t)&=\sum_{k,p}\Big [f^{}_{R}(\epsilon^{}_{p})-f^{}_{L}(\epsilon^{}_{k})\Big ]\int_{-\infty}^{t}dt^{}_{1}e^{\eta t^{}_{1}}\\
&\times\Big (e^{i(\epsilon^{}_{k}-\epsilon^{}_{p})(t-t^{}_{1})}[W^{}_{LR}(t)W^{\dagger}_{LR}(t^{}_{1})]^{}_{\sigma'\sigma} \nonumber\\
&\hspace{5mm}+
e^{i(\epsilon^{}_{p}-\epsilon^{}_{k})(t-t^{}_{1})}[W^{}_{LR}(t^{}_{1})W^{\dagger}_{LR}(t)]^{}_{\sigma'\sigma}\Big )\ .
\nonumber
\end{align}


\section{Explicit expressions for $I_{L}$ and $\dot{\bf M}_{L}$ for a circularly-polarized field}
\label{AppB}

As discussed in the main text, the time and spin dependence of the tunneling amplitude $W_{LR}(t)$ results from the Aharonov-Casher
phase operator, in conjunction with an oscillating electric field.
For a circularly-polarized electric field
the tunneling amplitude is given by Eq. (\ref{W}), and the vector $\hat{\bf v}(t)$ is given by Eq. (\ref{v}). This yields Eq. (\ref{pro}).

It follows that the trace of the matrix on the right hand-side, which appears in the expression for the particle current (\ref{opIL}),
is
\begin{align}
\label{tr}
{\rm Tr}[
W^{}_{LR}(t)W^{\dagger}_{LR}(t^{}_{1})]= \\
&\hspace{-20mm}2|W^{}_{0}|^{2}\Big (\cos^{2}_{}(k^{}_{\rm so}d)+\sin^{2}_{}(k^{}_{\rm so}d)\cos[\Omega(t-t^{}_{1})]\Big )\ .
\nonumber
\end{align}
As it depends only upon the times' difference $t-t^{}_{1}$, the particle current through the junction does not vary with $t$.
The same type of time-dependence appears in the $\sigma_{x}$ component of the product in Eq. (\ref{pro}), which can be written as
\begin{align}
\label{sx}
{\rm Tr}[\sigma^{}_{x}
W^{}_{LR}(t)W^{\dagger}_{LR}(t^{}_{1})]= \\
&\hspace{-20mm}-
2i|W^{}_{0}|^{2}\sin^{2}_{}(k^{}_{\rm so}d)\sin[\Omega(t-t^{}_{1})]\ .
\nonumber
\end{align}
As a result, the  $x$ component of the spin current, $\dot{M}_{L,x}$, is also independent of time.
On the other hand,
 the spin currents $\dot{M}_{L,y}$ and $\dot{M}_{L,z}$,  are determined by
\begin{align}
\label{syz}
{\rm Tr}[\sigma^{}_{y}
W^{}_{LR}(t)W^{\dagger}_{LR}(t^{}_{1})]&= \\
&\hspace{-20mm}2i|W^{}_{0}|^{2}\cos^{}_{}(k^{}_{\rm so}d)\sin^{}_{}(k^{}_{\rm so}d)[\cos(\Omega t)-\cos(\Omega t^{}_{1})]\ ,
\nonumber\\
{\rm Tr}[\sigma^{}_{z}
W^{}_{LR}(t)W^{\dagger}_{LR}(t^{}_{1})]&= \nonumber\\
&\hspace{-20mm}2i|W^{}_{0}|^{2}\cos^{}_{}(k^{}_{\rm so}d)\sin^{}_{}(k^{}_{\rm so}d)[\sin(\Omega t)-\sin(\Omega t^{}_{1})]\ ,
\nonumber
\end{align}
and consequently are time dependent.

Using Eqs. (\ref{R}) and (\ref{tr}) in conjunction with Eq. (\ref{opIL}), one finds the particle current into the left lead,
\begin{widetext}
\begin{align}
I^{}_{L}&=2|W^{}_{0}|^{2}\sum_{k,p}\Big [f^{}_{R}(\epsilon^{}_{p})-f^{}_{L}(\epsilon^{}_{k})\Big ]
\int_{-\infty}^{t}dt^{}_{1}e^{\eta t^{}_{1}}
\Big (\cos^{2}_{}(k^{}_{\rm so}d)+\sin^{2}_{}(k^{}_{\rm so}d)\cos[\Omega(t-t^{}_{1})]\Big )
\Big (e^{i(\epsilon^{}_{k}-\epsilon^{}_{p})(t-t^{}_{1})}+
e^{i(\epsilon^{}_{p}-\epsilon^{}_{k})(t-t^{}_{1})}\Big )\nonumber\\
&=4\pi |W^{}_{0}|^{2}\sum_{k,p}\Big [f^{}_{R}(\epsilon^{}_{p})-f^{}_{L}(\epsilon^{}_{k})\Big ]
\Big (\cos^{2}_{}(k^{}_{\rm so}d)\delta(\epsilon^{}_{k}-\epsilon^{}_{p})+\frac{1}{2}
\sin^{2}_{}(k^{}_{\rm so}d)\Big [\delta(\epsilon^{}_{k}-\epsilon^{}_{p}+\Omega)+
\delta(\epsilon^{}_{k}-\epsilon^{}_{p}-\Omega)\Big ]\Big )\ .
\label{IL1}
\end{align}
\end{widetext}
To zeroth order in $\Omega$ the terms within the round brackets reduce to $\delta(\epsilon_k - \epsilon_p)$; hence up to linear order in $\Omega$ the conductance is not affected by the spin-orbit coupling. The particle current into the right terminal is obtained from Eq. (\ref{IL1}) upon interchanging $k$ with $p$ and $L$ with $R$. Hence $I_{R}=-I_{L}$.

Equations (\ref{R}) and (\ref{sx}) in conjunction with Eq. (\ref{opML}) give 
\begin{widetext}
\begin{align}
\dot{M}^{}_{L,x}&=
2|W^{}_{0}|^{2}\sin^{2}_{}(k^{}_{\rm so}d)\sum_{k,p}\Big [f^{}_{R}(\epsilon^{}_{p})-f^{}_{L}(\epsilon^{}_{k})\Big ]
\int_{-\infty}^{t}dt^{}_{1}e^{\eta t^{}_{1}}
\Big (-ie^{i(\epsilon^{}_{k}-\epsilon^{}_{p})(t-t^{}_{1})}+ie^{i(\epsilon^{}_{p}-\epsilon^{}_{k})(t-t^{}_{1})}\Big )\sin[(\Omega(t-t^{}_{1})]\nonumber\\
&
=2\pi |W^{}_{0}|^{2}\sin^{2}_{}(k^{}_{\rm so}d)\sum_{k,p}\Big [f^{}_{R}(\epsilon^{}_{p})-f^{}_{L}(\epsilon^{}_{k})\Big ]
\Big (\delta(\epsilon^{}_{k}-\epsilon^{}_{p}+\Omega)-\delta(\epsilon^{}_{k}-\epsilon^{}_{p}-\Omega)\Big )\ .
\label{dMx}
\end{align}
The  terms in the round brackets give a contribution of order $\Omega$ to  $\dot{M}^{}_{L,x}$.
The corresponding spin current in the right reservoir,
$\dot{M}^{}_{R,x}$, is obtained from Eq. (\ref{dMx}) by interchanging $k$ with $p$ and $L$ with $R$; as seen,  it has the same sign as that of
$\dot{M}^{}_{L,x}$.

From the first of Eqs. (\ref{syz}) in conjunction with Eqs. (\ref{opML}) and (\ref{R}) it follows that
\begin{align}
&\dot{M}^{}_{L,y}=
2|W^{}_{0}|^{2}\cos(k^{}_{\rm so}d)\sin(k^{}_{\rm so}d)\sum_{k,p}\Big [f^{}_{R}(\epsilon^{}_{p})-f^{}_{L}(\epsilon^{}_{k})\Big ]\nonumber\\
&\times\int_{-\infty}^{t}dt^{}_{1}e^{\eta t^{}_{1}}
\Big (ie^{i(\epsilon^{}_{k}-\epsilon^{}_{p})(t-t^{}_{1})}-ie^{i(\epsilon^{}_{p}-\epsilon^{}_{k})(t-t^{}_{1})}\Big )\Big [\cos(\Omega t)-\cos(\Omega t^{}_{1})\Big ]\ .
\end{align}
As
\begin{align}
\cos(\Omega t)-\cos(\Omega t^{}_{1})=\cos(\Omega t)
-\frac{1}{2}\Big (\cos(\Omega t)\cos [\Omega (t-t^{}_{1})]+\sin(\Omega t)\sin [\Omega (t-t^{}_{1})]\Big )\ ,
\end{align}
we find
\begin{align}
&\dot{M}^{}_{L,y}=
2|W^{}_{0}|^{2}\sin(2k^{}_{\rm so}d)
\sum_{k,p}\Big [f^{}_{R}(\epsilon^{}_{p})-f^{}_{L}(\epsilon^{}_{k})\Big ]\nonumber\\
&\times\Big \{\cos(\Omega t)\Big (\frac{\cal P}{\epsilon^{}_{k}-\epsilon^{}_{p}}-\frac{1}{2}\Big [\frac{\cal P}{\epsilon^{}_{k}-\epsilon^{}_{p}+\Omega}+
\frac{\cal P}{\epsilon^{}_{k}-\epsilon^{}_{p}-\Omega}\Big ]\Big )+
\pi \sin(\Omega t)
\Big (\delta(\epsilon^{}_{k}-\epsilon^{}_{p}+\Omega)-
\delta(\epsilon^{}_{k}-\epsilon^{}_{p}-\Omega)\Big )\Big \}
\ ,
\label{Sy}
\end{align}
where ${\cal P}$ indicates the principal part.
Likewise,
\begin{align}
\dot{M}^{}_{L,z}&=
2|W^{}_{0}|^{2}\cos(k^{}_{\rm so}d)\sin(k^{}_{\rm so}d)\sum_{k,p}\Big [f^{}_{R}(\epsilon^{}_{p})-f^{}_{L}(\epsilon^{}_{k})\Big ]\nonumber\\
&\times\int_{-\infty}^{t}dt^{}_{1}e^{\eta t^{}_{1}}
\Big (ie^{i(\epsilon^{}_{k}-\epsilon^{}_{p})(t-t^{}_{1})}-ie^{i(\epsilon^{}_{p}-\epsilon^{}_{k})(t-t^{}_{1})}\Big )\Big [\sin(\Omega t)-\sin(\Omega t^{}_{1})\Big ]\ ,
\end{align}
with
\begin{align}
\sin(\Omega t)-\sin(\Omega t^{}_{1})=\sin(\Omega t)
-\frac{1}{2}\Big (\sin(\Omega t)\cos [\Omega (t-t^{}_{1})]-\cos(\Omega t)\sin [\Omega (t-t^{}_{1})]\Big )\ ,
\end{align}
leads to
\begin{align}
&\dot{M}^{}_{L,z}=2|W^{}_{0}|^{2}\sin(2k^{}_{\rm so}d)
\sum_{k,p}\Big [f^{}_{R}(\epsilon^{}_{p})-f^{}_{L}(\epsilon^{}_{k})\Big ]\nonumber\\
&\times\Big \{
\sin(\Omega t)\Big (\frac{\cal P}{\epsilon^{}_{k}-\epsilon^{}_{p}}-\frac{1}{2}\Big [\frac{\cal P}{\epsilon^{}_{k}-\epsilon^{}_{p}+\Omega}+
\frac{\cal P}{\epsilon^{}_{k}-\epsilon^{}_{p}-\Omega}\Big ]\Big )-\pi
\cos(\Omega t)
\Big (\delta(\epsilon^{}_{k}-\epsilon^{}_{p}+\Omega)-
\delta(\epsilon^{}_{k}-\epsilon^{}_{p}-\Omega)\Big )\Big \}\ .
\label{Sz}
\end{align}
The principal parts in Eqs. (\ref{Sy}) and (\ref{Sz}) are of order $\Omega^{2}$, and hence can be neglected in the small $\Omega$ limit. The remaining terms, which are of order $\Omega$, rotate in the $y-z$ plane, along  the vector $[0,\sin(\Omega t),-\cos(\Omega t)]$.
Remarkably enough,  the time integral over this vector corresponds to a transverse magnetization  parallel to the effective  magnetic field, which is directed along $\hat{\bf v}(t)$, Eq. (\ref{v}).
\end{widetext}


\begin{thebibliography}{99}
\bibitem{com2}
Alternatively, the fields can be generated by  a single back gate  and two other gates placed very close to the sides of the wire as in Ref.~\onlinecite{nygard}.
\bibitem{com1}
The final results are expressed as traces over the $\sigma$'s, and therefore they do not depend on the quantization axes which define the indices $\sigma,~\sigma'$.
\bibitem{mjcomment}
If the densities of states in the terminals are energy dependent (and equal),
${\cal N}_L(\epsilon)= {\cal N}_R(\epsilon)={\cal N}(\epsilon)$, $G$ in Eq.~(\ref{ILresult})
should be multiplied by
$\{1 + \sin^2(k_{\rm so}d)[{\cal N}(\epsilon_{\rm F} + \Omega) + {\cal N}(\epsilon_{\rm F} - \Omega) - 2{\cal N}(\epsilon_{\rm F})]/(2{\cal N}(\epsilon_{\rm F}))\}$. Hence,
there is a small SOI effect on the particle current of order $(\Omega/\epsilon_{\rm F})^2$.
\bibitem{circuits}
In a closed circuit, the two terminals are connected to each other via a voltage source (a battery), which
generates a bias voltage, causing the current flow through the circuit. When there is no bias voltage, the two terminals can also be grounded.
 In an open circuit, the weak-link device and
 the terminals are electrically isolated from the
environment. The term ``open/closed circuit" (in electrical
engineering) should not be confused with ``open/closed system"
in physics: the ``open circuit" corresponds to the ``closed system" and vice versa.
\end{thebibliography}
\end{document}